\documentclass[%
superscriptaddress,
 amsmath,amssymb,
 aps,
 prstab, twocolumn, longbibliography,
]{revtex4-1}

\usepackage{tabularx}
\usepackage{textcomp}
\usepackage{graphicx}%
\usepackage[dvipsnames]{xcolor}
\usepackage{chemmacros}
\usepackage{amsmath}
\usepackage{amssymb}
\usepackage{dcolumn}
\usepackage{graphicx}
\usepackage{longtable}
\usepackage{bm}
\usepackage{color}
\usepackage{epsfig}
\usepackage{multirow}

\usepackage[colorlinks, citecolor=blue, urlcolor=black, bookmarks=false, linkcolor=blue, hypertexnames=true]{hyperref}

\begin{document}


\title{Polarization selectivity of aloof-beam electron energy-loss spectroscopy in one-dimensional ZnO nanorods}

\author{Yao-Wen Yeh}
\email{yaowen@physics.rutgers.edu}
\affiliation{Department of Physics and Astronomy, Rutgers University, Piscataway, New Jersey 08854, USA}

\author{Sobhit Singh}
\email{sobhit.singh@rutgers.edu}
\affiliation{Department of Physics and Astronomy, Rutgers University, Piscataway, New Jersey 08854, USA}

\author{David Vanderbilt}
\affiliation{Department of Physics and Astronomy, Rutgers University, Piscataway, New Jersey 08854, USA}

\author{Philip E. Batson}
\affiliation{Department of Physics and Astronomy, Rutgers University, Piscataway, New Jersey 08854, USA}


\begin{abstract}
Orientation dependent electronic properties of wurtzite zinc oxide nanorods are characterized by aloof beam electron energy-loss spectroscopy (EELS) carried out in a scanning transmission electron microscope (STEM). The two key crystal orientation differentiating transitions specific to the in-plane (13.0 eV) and out-of-plane (11.2 eV) directions with respect to the wurtzite structure are examined by first principles density-functional theory calculations. 
We note some degree of orientation dependence at the onset of direct band gap transition near 3.4 eV. 
We demonstrate that good polarization selectivity can be achieved by placing the electron probe at different locations around the specimen with increasing impact parameter while keeping the beam-specimen orientation fixed. The observed results are qualitatively elucidated in terms of the perpendicular electric fields generated by the fast electron (60 kV) used in the microscope. 
The fact that good polarization selectivity can be achieved by aloof beam EELS without the requirement of sample reorientation is an attractive aspect from the characterization method point of view in the STEM-EELS community. 
\end{abstract}


\maketitle

\section{Introduction}
\label{sec:intro}
Scanning transmission electron microscope coupled electron energy-loss spectroscopy (STEM-EELS) plays an important role in the development and characterization of nanomaterials due to its high spatial resolution in both imaging and spectroscopy. In the more straightforward intersecting beam-specimen configuration where the interaction is dominated by impact scattering, high resolution STEM-EELS is able to probe local material properties such as interface states~\cite{Batson1993,Muller1999} and single-atom dopant electronic properties~\cite{Ramasse2013}, and even vibrational properties~\cite{Lagos2018,Venkatraman2019,Hage2020} aided by recent development in energy monochromators~\cite{Krivanek2014}. Typically, these high spatial resolution STEM-EELS experiments are carried out with a focused electron probe with a large convergence semi-angle ($\alpha$), which allows the incident electrons to excite the specimen with significant momentum transfer. With a common circular collection aperture used in the on-axis arrangement, the acquired transmitted signals lead to the so called momentum integrated energy-loss spectra~\cite{Venkatraman2019, Rafferty1998}. When the specimen of interest has complex electronic and phonon structures in the reciprocal space, the resulting EELS spectra would also reflect the structural complexity, which can make it difficult to fully interpret the observed spectra without prior knowledge of the material. 

Meanwhile, in the aloof beam (non-intersecting probe-specimen) configuration where the interaction is dominated by dipole scattering, the electron beam behaves as a near-field spectroscopy probe~\cite{Cohen1998} and can be used to study long range electric field ($\bf{E}$) related phenomena such as vibrational properties~\cite{Radtke2017,Radtke2019}, surface plasmon-polaritons in metal and doped-semiconductor nanostructures~\cite{Batson1982,Rossouw2013,Yang2020}, and surface phonon-polaritons in insulating nanostructures~\cite{Lagos2017,Qi2019}. Recently, Rez \textit{et al.}~\cite{Rez2016} and Crozier~\cite{Crozier2017} discussed the non-destructive nature of the aloof beam EELS and expanded its applications in studying vibrational and electronic properties for electron beam sensitive materials. In addition to the non-destructive characterization, Radtke \textit{et al.}~\cite{Radtke2019} have further applied aloof beam EELS to study polarization dependent vibrational properties of anisotropic materials demonstrating that good polarization selectivity can be achieved by aloof beam EELS. We note that orientation dependent anisotropic properties have traditionally been measured in the beam-specimen intersecting configuration with small beam convergence ($\alpha<0.5$ mrad) in the past.~\cite{Leapman1979,Leapman1983,Arenal2007}

Here, we explore the potential application of aloof beam EELS to characterize the polarization-dependent electronic structures of uniaxial crystals in scenarios where specimen reorientation is challenging to achieve by the microscope goniometer. 
In particular, we study one dimensional wurtzite zinc oxide (w-ZnO) nanorods in a fixed beam-to-specimen orientation, compare the results obtained with the intersecting and aloof beam configurations, and demonstrate that the aloof beam EELS offers good polarization selectivity when the impact parameter, {\it i.e.,} specimen-to-probe distance, is large enough ($\sim$6\,nm). The prominent anisotropic features that differentiate the in-plane ($\bot$ to $c$ axis) and out-of-plane ($\parallel$ to $c$ axis) responses of w-ZnO nanorods are examined by first-principles density-functional theory (DFT) calculations. 
Our results assert that the previously debated transition at 11.2 eV in w-ZnO is not due to a surface plasmon~\cite{Wang2005,Wang2007}, but instead corresponds to an orientation-dependent interband transition from the occupied Zn-$3d$ states to the unoccupied O-$2p_{z}$ states near the zone center~\cite{Huang2011}.  
Finally, we discuss the aloof beam results in terms of the electric fields associated with the excitation source (incident electrons) rather than the dielectric function that describes the response to an external perturbation. This approach provides qualitative but intuitive understanding of the results. Moreover, the fact that good polarization selectivity can be achieved by aloof beam EELS without specimen reorientation is another attractive aspect to the technique, as many technologically important materials can be anisotropic and have large aspect ratios~\cite{Xia2003,Sajanlal2011}.

\section{Materials and methods}
\label{sec:method}
w-ZnO nanorods were synthesized via the hydrothermal route described by Cheng and Samulski~\cite{Cheng2004}. The two starting solutions were obtained by mixing (1) 0.1M of zinc acetate dihydrate (\ch{Zn(Ac)2 * 2 H2O}) in methanol and (2) 0.5M of sodium hydroxide (\ch{NaCl}) in ethanol. 60 ml of of the mixed solution (1:1 volume ratio) was then transferred to a Teflon-lined stainless steel autoclave and heated at 150 $^{\circ}$C for 8 hours. After the autoclave was cooled to room temperature, the white precipitates containing ZnO nanorods were washed with de-ionized water four times and dried at 60 $^{\circ}$C in an oven overnight. The dried powder was then annealed in air at 400 $^{\circ}$C for 1 hour to improve its crystallinity. The electron microscopy sample was prepared by first dispersing the annealed ZnO powder in ethanol via sonication and the sample containing solution was then drop cast onto a lacey carbon coated copper grid.

\begin{figure}
    \centering
    \includegraphics[width=0.45\textwidth]{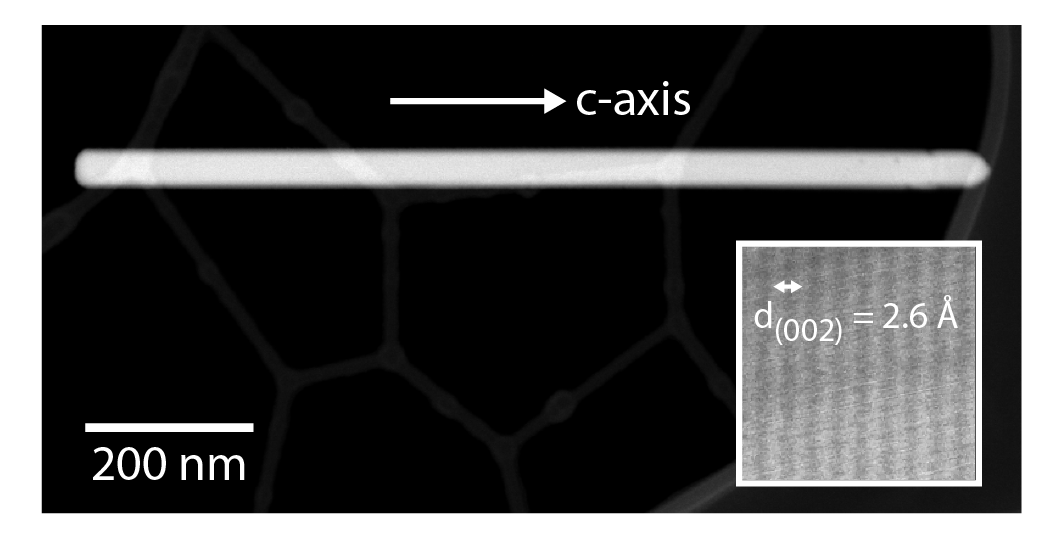}
    \caption{
    Annular dark-field STEM image of a single suspended 1100 by 40 nm ZnO nanorod on lacey carbon support. The long axis of the nanorod corresponds to the $c$ axis of w-ZnO. The inset shows a high resolution image of (002) lattice fringes with the spacing of 2.6 \r{A}.}
    \label{fig:HAADF}
\end{figure}

The synthesized nanorods have lengths that range from 90 to 1200 nm and widths that range from 35 to 40 nm.  A typical synthesized long nanorod is shown in Fig. \ref{fig:HAADF}. As reported by Cheng and Samulski~\cite{Cheng2004}, the rods preferentially grow along the $c$ axis. Regardless of the rod lengths, the $c$ axis of w-ZnO is always along the rod axis as confirmed by the high resolution image shown in the inset of Fig. \ref{fig:HAADF}, where $d_{002}=2.6$ \r{A} was observed.

The STEM-EELS experiments reported here were carried out in a monochromated Nion UltraSTEM operating at 60 kV. The beam convergence and EELS collection semi-angles were set at 30 and 16 mrad, respectively. The energy resolution for the spectra is 87.2\,meV according to the full-width-half-maximum of the zero-loss peak.

DFT calculations were performed using the Projector Augmented Wave (PAW) method as implemented in the Vienna Ab initio Simulation Package (VASP)~\cite{Kresse96a, Kresse96b, KressePAW}. Twelve and six valence electrons were considered in the zinc and oxygen PAW pseudopotentials, respectively. The Perdew-Burke-Ernzerhof generalized-gradient approximation for solids (GGA-PBEsol) was used to compute the exchange-correlation functional~\cite{PBEsol}. 
The reciprocal space was sampled using a $\Gamma$-centered k-mesh of size 16$\times$16$\times$10 together with a kinetic energy cutoff of 650\,eV for plane waves. 
The energy convergence and force convergence criteria were set to $10^{-8}$ eV and $10^{-4}$ eV/\AA, respectively. 

All the DFT calculations were carried out in a four-atom hexagonal unit cell of w-ZnO (space group: $P6_{3}mc$). The GGA-PBEsol optimized lattice parameters are $a=b=3.236$\,\AA, and $c=5.223$\,\AA. which are in good agreement with the experimental data~\cite{Adachi1999, Wang_JPCM2004}. The employed DFT predicts w-ZnO to be a direct band gap semiconductor with an energy band gap of 0.7 eV at $\Gamma$. We apply a scissor operation of 2.7 eV to obtain the reported experimental bandgap of 3.4 eV in w-ZnO~\cite{Adachi1999, Wang_JPCM2004, FRANKLIN2013, NEJATIPOUR2015}.  
The calculated electronic bandstructure (in particular, the structure of the conduction bands and valence bands near the Fermi level) matches reasonably well with the previously reported hybrid and quasi-particle $GW$ calculations~\cite{Preston2008, SchleifePRB2009, Schleife2009, GoriPRB2010, OchiPRL2017, AtaidePRB2017, PrestonPRB2011, StankovskiPRB2011, ZhangPRB2018}.
The frequency-dependent imaginary dielectric function $\epsilon_{2}\,(\omega)$ was calculated using the method described in Ref.~\cite{GajdoPRB2006}. The convergence of $\epsilon_{2}\,(\omega)$ was ensured by increasing the number of empty conduction bands. 
The calculated $\epsilon_{2}\,(\omega)$ is in good agreement with the previous studies~\cite{FritschIEEE2005, Preston2008, Akiyama_2008, fritsch2008intensity, schmidt2008vacuum, SchleifePRB2009, GoriPRB2010} and gives a good description of the recorded EELS spectra in this work.

\section{Results and discussion}
\label{sec:result}
w-ZnO is a uniaxial material of hexagonal structure and it possesses two sets of dielectric properties - one associated to the in-plane direction ($a$ and $b$ axes) and the other associated to the out-of-plane ($c$ axis) direction. The oxygen $2p_{xy}$ (in-plane) and $2p_z$ (out-of-plane) orbitals play an important role in governing the anisotropic electronic properties of w-ZnO. Our DFT results, presented below, as well as previous studies~\cite{Adachi1999,Wang_2004,Ding2005, FritschIEEE2005, Zhang2006, Preston2008, Akiyama_2008, fritsch2008intensity, schmidt2008vacuum, SchleifePRB2009, GoriPRB2010}, suggest that there are several distinct optical (dipole-dipole) interband transitions in w-ZnO at energies of $\sim$3.4, 9.1, 11.2, 13.0, 13.5, and 14.8 eV, as listed in Table~\ref{table:transitions}. Although the transitions at $\sim$3.4, 9.1, and 13.5 eV are well documented in the existing literature~\cite{Adachi1999,Wang_2004,Ding2005, FritschIEEE2005, Zhang2006, Preston2008, Akiyama_2008, fritsch2008intensity, schmidt2008vacuum,  SchleifePRB2009, GoriPRB2010}, less is known about the other listed  transitions. Notably, the transitions at 9.1, 13.5, and 14.8 eV are not very sensitive to the polarization orientation, as these are common to both the in-plane and out-of-plane polarization orientations of the $\bf{E}$-field. In contrast, the transitions at 3.4, 3.5, 11.2, and 13.0 eV exhibit strong orientation dependence due to the anisotropic nature of the O-$2p_{xy}$ and O-$2p_z$ orbitals in w-ZnO.

\begin{table}[h!]
\caption{Calculated optical transitions below 15.0 eV,  from the occupied to unoccupied states near the Fermi level, and their polarization orientation with respect to the $c$ axis in w-ZnO \\ }
\centering
\renewcommand{\arraystretch}{1.5}
 \begin{tabular}{||c c c ||} 
 \hline
 Energy (eV) & Orientation & Assignment \\ 
 \hline\hline
  3.4  & $ \bot $ & O-$2p_{xy}$ $\rightarrow$ Zn-$3s$ \\
  3.5  & $ \parallel$ & O-$2p_{z}$ $\rightarrow$ Zn-$3s$ \\
 9.1  & $\bot \& \parallel$ & Zn-$3d$ $\rightarrow$ O-$2p$ \\ 
 11.2 & $\parallel$ & Zn-$3d$ $\rightarrow$ O-$2p_{z}$ \\
 13.0 & $\bot$ & Zn-$3d$ $\rightarrow$ O-$2p_{xy}$ \\
 13.5 & $\bot \& \parallel$ & O-$2p$  $\rightarrow$ Zn-$3s$  \\
 \multirow{2}{*}{14.8} & $\bot \& \parallel$ & Zn-$3d$ $\rightarrow$ O-$2p$ \\
                        & $\bot \& \parallel$ & O-$2p$ $\rightarrow$ \{Zn-$3d$, Zn-$3s$\} \\

 \hline
 \end{tabular}
\label{table:transitions}
\end{table}

\begin{figure}
    \centering
    \includegraphics[width=0.45\textwidth]{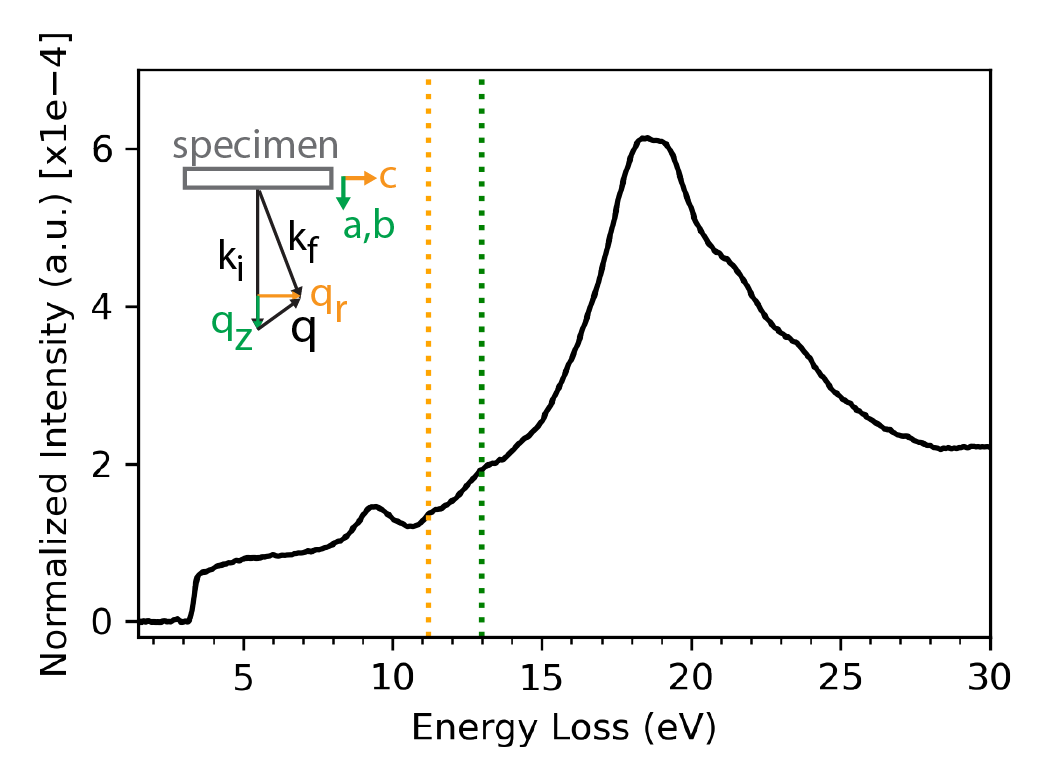}
    \caption{
    Valence EELS spectrum acquired with the beam intersecting configuration after Fourier-log deconvolution~\cite{Egerton2011} and background removal. Three major peaks are the onset of the band gap around 3.4 eV, Zn-$3d$ to O-$2p$ transition around 9.1 eV, and the bulk plasmon at 18.8 eV. While not being immediately obvious, the orientation specific features can be seen around 11.2 eV (out-of-plane) and 13.0 eV (in-plane) as guided by the dotted orange and green lines, respectively. The inset shows the scattering geometry with $\hat{c}\parallel\hat{q_r}$ and $\hat{a}$ and $\hat{b}\parallel\hat{q_z}$
    }
    \label{fig:BulkSpectrum}
\end{figure}

In our experiments, the ZnO nanorod is oriented such that the incident electron beam is perpendicular to the crystal $c$ axis. Fig.~\ref{fig:BulkSpectrum} shows the EELS spectrum of ZnO obtained using the specimen-intersecting beam and the inset shows the scattering geometry. The three most apparent spectral features are the direct band gap near 3.4 eV, Zn-$3d$ to O-$2p$ transition around 9.1 eV, and the bulk plasmon at 18.8 eV, which are consistent with the previous observations~\cite{Wang2005,Wang2007,Zhang2006,Ding2005,Huang2011}. With the large beam convergent angle ($\alpha=30$ mrad) used in the present study, the beam is able to excite transitions with polarization along both the parallel ($a$ and $b$ axes) and perpendicular ($c$ axis) to the beam (crystal) directions.

In conjunction with the large incident beam convergence angle, the moderate EELS collection angle ($\beta=16$ mrad) used in the present study leads to momentum integrated EELS~\cite{Rafferty1998}. Consequently, the orientation-specific 11.2 ($\parallel \,c$) and 13.0 eV ($\bot\,c$) transitions are always observed in the spectrum regardless of where the beam is placed, as long as the beam intersects the sample. 
However, the poor polarization selectivity observed in the intersecting configuration makes it impractical in studying orientation specific features. We should note that the scattered signals being collected can be further restricted to the beam parallel direction and achieve decent polarization selectivity in the intersection configuration by using a smaller aperture to reduce the collection angle~\cite{Batson2008}. However, such small aperture is not available in our current microscope setup. In addition, restricting momentum transfer causes unavoidable uncertainty-principle driven spread in the area that is sampled, degrading the desired spatial resolution of the measurement using the STEM. 

In addition, due to the considerable momentum transfer ($\vec{q}$ = $\vec{q_r}+\vec{q_z}$) in the intersecting scattering configuration, the transitions around 11.2, 13.0, 13.5, and 14.8 eV appear as minor humps which are partly obscured by the nearby transitions and the bulk plasmon. Moreover, the 11.2 eV hump was misinterpreted as a surface plasmon in the past~\cite{Wang2005,Wang2007}. In the present study, we focus on the anisotropic features of w-ZnO, and for more detailed spectral feature assignments we refer the readers to Refs.~\cite{Zhang2006, Huang2011}.

\begin{figure*}
    \centering
    \includegraphics[width=1.0\textwidth]{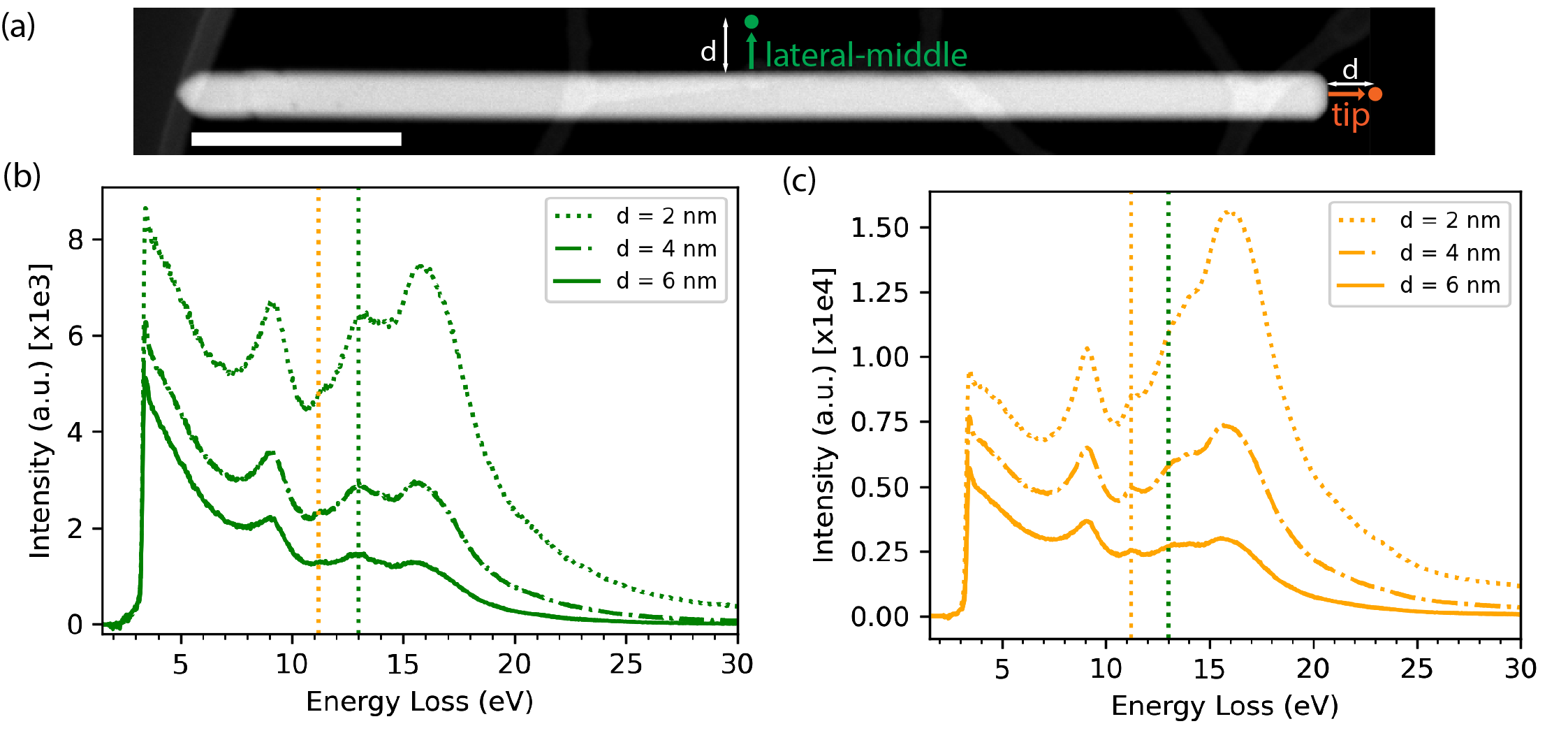}
    \caption{
    (a) The high annular dark field image of a 1100 by 40 nm ZnO nanorod (scale bar: 200 nm) where the green and orange dots indicate the electron beam locations where the aloof beam spectra were acquired. Three spectra were acquired at impact parameters of 2, 4, and 6 nm (b) along the green arrow near the lateral-middle position of the rod and (c) along the orange arrow near the rod tip. The dotted vertical line at 11.2 and 13.0 eV indicate the key anisotropic transitions that are specific to the direction parallel and perpendicular to the $c$ axis, respectively.}
    \label{fig:main}
\end{figure*}

To explore the polarization selectivity of the aloof beam EELS while keeping the rod orientation fixed, we acquired two sets of data points - one in the vicinity of the rod tip and the other in the vicinity of the lateral-middle position of the rod. Each set contains spectra acquired at three different impact parameters ($d$) of 2, 4, and 6 nm along the directions indicated by the orange and green arrows, as shown in Fig.~\ref{fig:main}(a). The acquired spectra near the lateral-middle and the tip of the rod are shown in Figs.~\ref{fig:main}(b) and \ref{fig:main}(c), respectively. Consistent with the calculated transitions listed in Table~\ref{table:transitions}, the not very orientation-specific transitions around 3.4 eV and the 9.1 and 13.5 eV peaks are observed in both aloof beam data sets, and the broad peak around 16.0 eV is the surface plasmon~\cite{Huang2011}. We note that the last non-orientation specific transition at 14.8 eV is overshadowed by the broad surface plasmon peak, making it difficult to identify. 

At small impact parameter ({\it i.e.,} $d$ = 2\,nm), both the 11.2 eV and 13.0 eV transitions are visible in both sets. However, as the impact parameter increases, the 11.2 eV and 13.0 eV transitions remain significant only when the electron beam is placed near the tip and lateral-middle position of the rod, respectively. Therefore, good polarization selectivity is achieved when $d$ = 6\,nm, despite the fact that the beam-parallel electric field ($\bot$ to the specimen $c$ axis) is always present. 
Unlike in Ref.~\cite{Radtke2019}, where the polarization-dependent spectra were acquired with two different beam-to-specimen orientations, the aloof beam spectra presented here were acquired with the same beam-to-specimen orientation but at different locations and different values of impact parameter.

To understand the observed aloof beam EELS results, we turn our attention to the electric fields generated by the incident electron, since the material response (induced dipole field) also scales with the external perturbation magnitude. The fast electrons (60 keV) used here can be considered as a broadband electromagnetic source~\cite{Garcia2010,MacielEscudero2019} producing electric fields in the frequency domain that can be broken down into two parts - one being perpendicular to the incident electron beam direction $E_{R}$ and the other being parallel to the beam $E_{z}$~\cite{Garcia2010,MacielEscudero2019,Masiello2012} as follows

\begin{equation}
\label{eq.E_perp}
E_{R}(x, y, z, \omega )=-\frac{2e \omega}{Rv^{2}\gamma } e^{i\omega z/v}K_{1}(\frac{\omega R}{v\gamma })[\,(x-d)\,\vec{x}+y\,\vec{y}\,]~, 
\end{equation}
\begin{equation}
\label{eq.E_para}
\text{and}~~ E_{z}(x, y, z,\omega )=i\frac{2e \omega}{Rv^{2}\gamma^{2} } e^{i\omega z/v}K_{0}(\frac{\omega R}{v\gamma })\,\vec{z}~.
\end{equation}

Here ($x, y, z$) is the observation location in the lab frame, $e$ is the elementary charge, $\omega$ is the angular frequency, $d$ is the impact parameter, $K_n$ is the modified Bessel function of second kind with index $n$, $R^2 = (x-d)^2 + y^2$, $v$ is the electron speed,  $\gamma = [1-(v/c)^2]^{-1/2}$ is the Lorentz factor, and $c$ is the speed of light in vacuum. On the other hand, we would also like to point out that the induced dipole field exerts forces on the incident electrons and wraps around the nanorod, occupying a three dimensional volume as illustrated in Ref. \cite{Batson2008}. The induced field components perpendicular and parallel to the incident electrons are responsible for the electron deflection and energy loss, respectively.

\begin{figure*}
    \centering
    \includegraphics[width=1.0\textwidth]{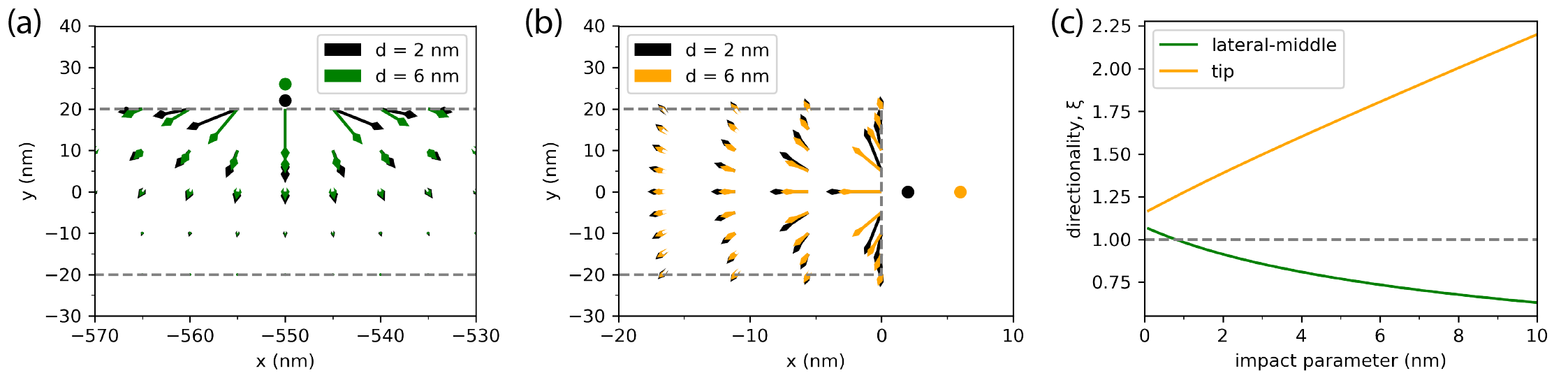}
    \caption{
    Electric field distributions, $E_R(R)$ generated by the fast electron at 11.2 eV within the nanorod when the aloof beam is placed near (a) the lateral-middle and (b) the tip of the rod. The black and green (orange)  circles represent the electron beam positions at the impact parameters of 2 and 6 nm, respectively. The nanorod area is delineated by the grey dashed lines. The fields are more 'collimated' toward the beam location as the impact parameter increases from 2 to 6 nm. The direction of the field arrows is reversed to avoid overlapping with the circles. (c) The directionality parameter versus the impact parameter near the lateral-middle (green) and the tip (orange) of the rod.}
    \label{fig:field}
\end{figure*}

We now focus our attention on the field distributions in the plane perpendicular to the beam direction (Eq.~\ref{eq.E_perp}), as this component is larger than the parallel one~\cite{Rafferty1998,Rez2016,Radtke2019} and contributes more to the overall electron-specimen interactions in the valence energy range considered here. As shown in Figs.~\ref{fig:field}(a) and \ref{fig:field}(b), we plot the perpendicular electric field distributions (Eq.~\ref{eq.E_perp}) at 11.2 eV within the nanorod area (delineated by the grey dashed line) when the beam is placed outside the lateral-middle and the tip of the rod, respectively. The $x$ and $y$ coordinates used here are referenced to the long ($c$ axis) and short axes of the ZnO crystal, and we refer the electric field components parallel and perpendicular to the $c$ axis as $E_{\parallel}$ and $E_{\bot}$, respectively. It can be seen that the resultant field becomes more directional toward the beam location when the beam is placed further away from the specimen. 

To visualize how the resultant field becomes more `collimated' in the plane perpendicular to the beam direction as the impact parameter increases, we define a directionality parameter $\xi=\Sigma(E_\parallel)/\Sigma(E_\bot)$, where $\Sigma(E_\parallel)$ and $\Sigma(E_\bot)$ are sums over a representative sampling of points within the nanorod. 
Figure~\ref{fig:field}(c) shows the variation of $\xi$ as a function of the impact parameter $d$. It is found that the electron beam generates electric fields that are more parallel to the $c$ axis near the tip (orange curve) than the lateral-middle (green curve) position of the rod. 
To put it more generally, when the specimen is placed on one side next to the beam with the extent no larger than its corresponding half-space, the fields that are generated by the beam and felt by the specimen are inherently asymmetric and directional in the aloof beam configuration. The larger the impact parameter, the more the aloof beam EELS resembles polarized optical spectroscopy. Therefore, in the case of w-ZnO, the 11.2 eV feature is only noticeable at large impact parameter near the tip but not at the lateral-middle position of the rod. A similar argument can be extended for the 13.0 eV feature.

\begin{figure}[htb!]
    \centering
    \includegraphics[width=0.47\textwidth]{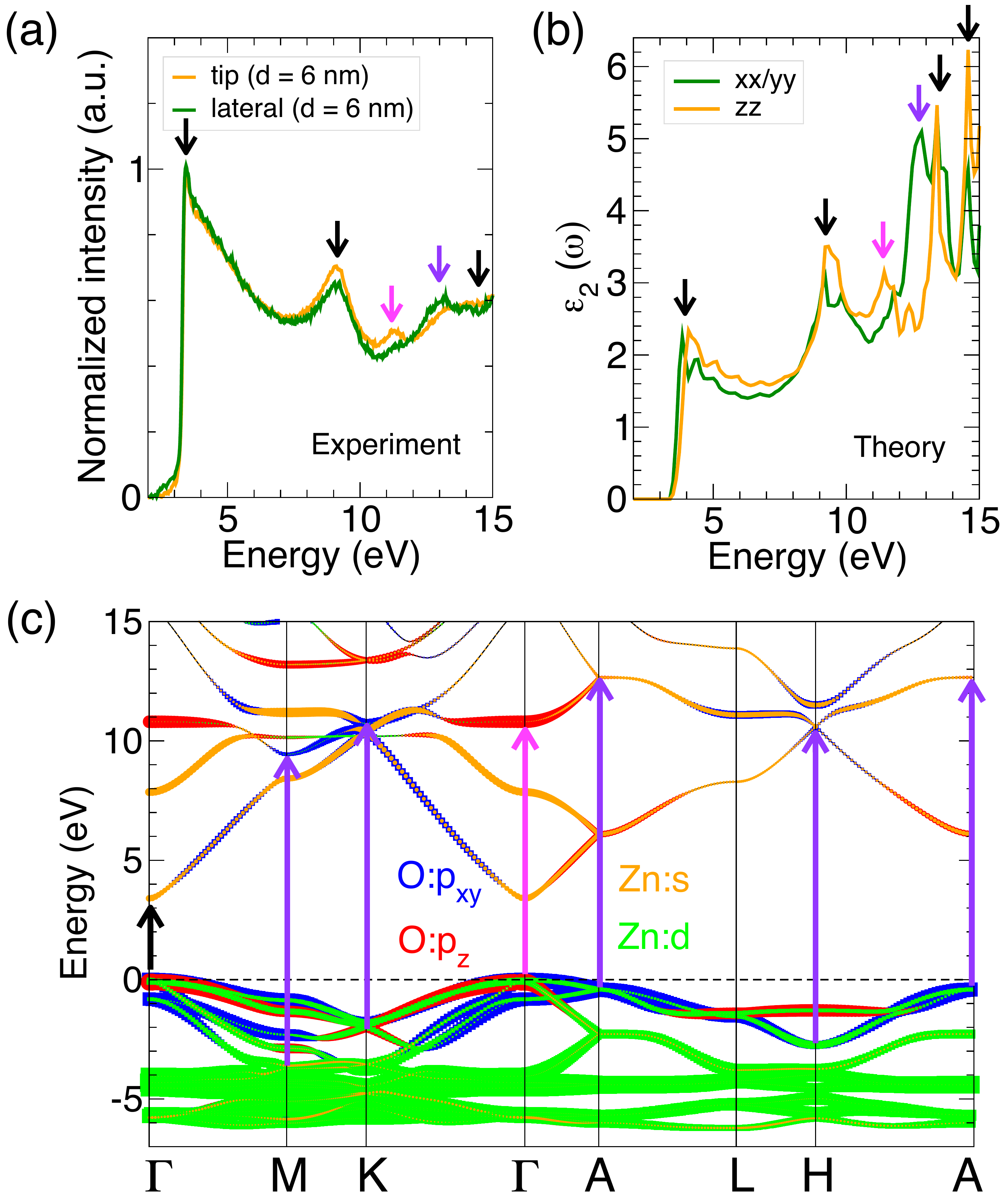}
    \caption{
    (a) Normalized aloof beam spectra ($d$ = 6 nm) acquired near the tip (orange) and the lateral-middle position (green) of the rod. The two anisotropic features at 11.2 eV ($E\parallel c$) and around 13.0 eV ($E\bot c$), marked using magenta and purple arrows, are differentiated near the tip and the lateral-middle position of the rod, respectively. The observed spectra demonstrate good polarization selectivity of the aloof beam EELS. (b) Polarization-orientation dependent imaginary dielectric functions, $\epsilon _2(\omega)$, calculated along xx/yy ($E\bot c$) and zz ($E\parallel c$) directions. The arrows mark the transitions of interest. (c) DFT calculated orbital-resolved electronic band structure of w-ZnO. Magenta and purple arrows depict the observed  inter-band optical transitions at 11.2 eV ($E\parallel c$) and near 13.0 eV ($E\bot c$), whereas black arrow denotes the direct band gap transition at 3.4 eV.  } 
    \label{fig:exp_theory}
\end{figure}

To aid the visual comparison of the observed aloof spectra, we scale the spectra with respect to the Bessel function and frequency parts of Eq.~\ref{eq.E_perp} and normalize the intensity around the band gap as the dielectric functions around the gap have similar values. As shown in Fig.~\ref{fig:exp_theory}(a), the normalization facilitates visualization of the orientation-specific transitions at 11.2 eV ($E\parallel c$) and 13.0 eV ($E\bot c$) near the tip (orange) and the lateral-middle (green) positions, respectively. 

In order to pin down the nature of the observed optical transitions, we employ DFT calculations as detailed in the methods section. The calculated polarization-orientation dependent imaginary dielectric function $\epsilon_{2}(\omega)$, which contains full details of the optical-transition matrix elements and the interband absorption processes, is in remarkable agreement with the recorded EELS spectra for both orientations, as shown in Fig.~\ref{fig:exp_theory}(b). Magenta and purple arrows respectively mark the orientation specific $E\parallel c$ and $E\bot c$ transitions, respectively, whereas, the black arrows mark the transitions that are not strongly orientation sensitive. The orbital-resolved electronic bandstructure, shown in Fig.~\ref{fig:exp_theory}(c), further facilitates the identification of the observed interband transitions from occupied to unoccupied states, as discussed below. 

\begin{figure}[htb!]
    \centering
    \includegraphics[width=0.49\textwidth]{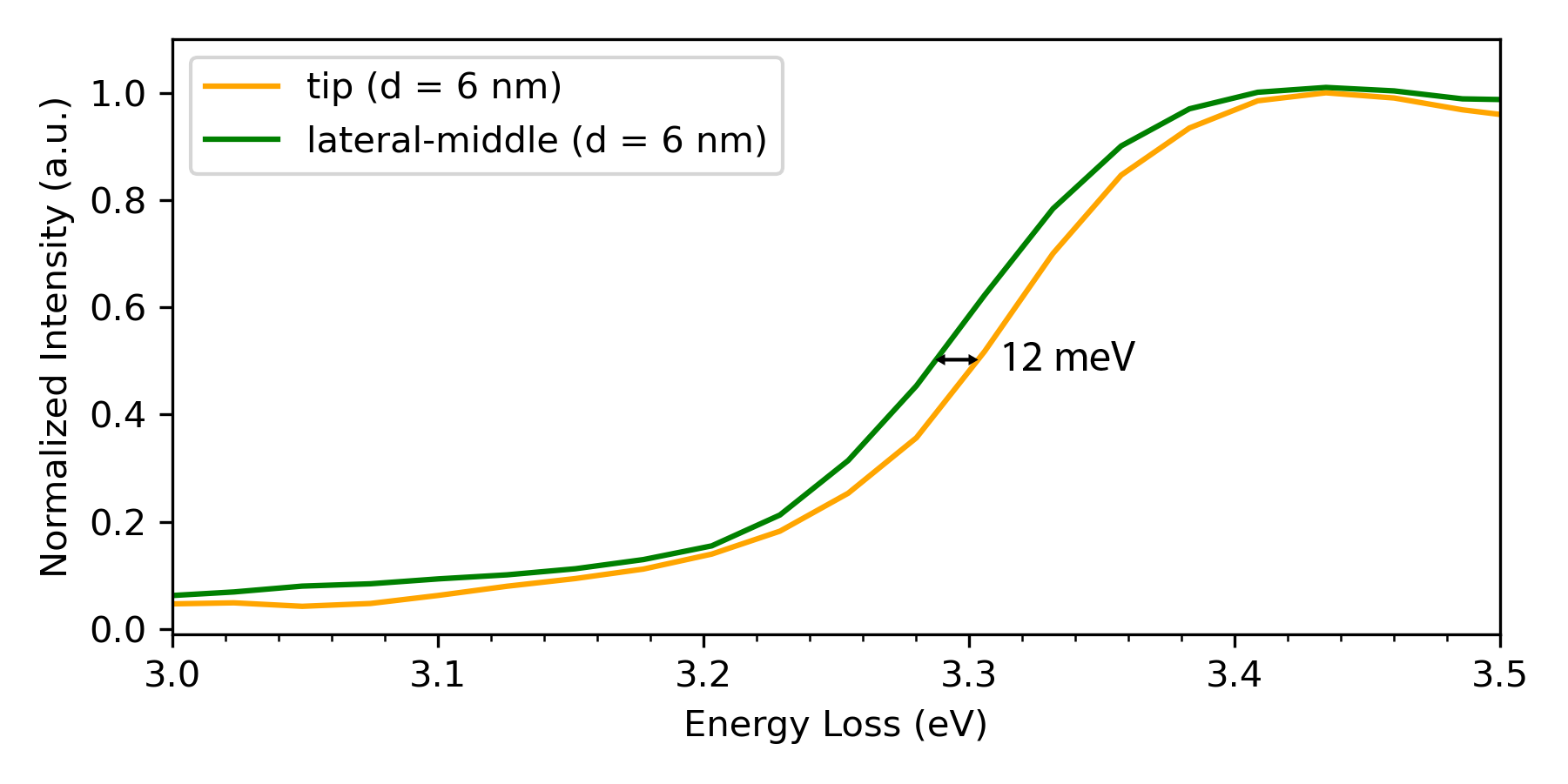}
    \caption{Band gap onset region of the aloof-beam spectra acquired at $d=6$ nm near the tip (orange) and lateral-middle (green) of the rod. Due to the mixed polarization signals presented in aloof-beam EELS, a minor onset difference of 12 meV is observed rather than the 70 meV difference observed in the fully polarized optical measurements reported in Ref.\cite{Adachi1999}.} 
    \label{fig:BandGap}
\end{figure}

We note that the onset of the direct band gap transition near 3.4 eV, although marked using a black arrow, is technically orientation dependent due to the splitting of the O-$2p$ states into a subset of O-$2p_{xy}$ and O-$2p_z$ states at the valence band maximum (see Fig.~\ref{fig:appendix_A}(b) in Appendix). This occurs due to the asymmetry of the wurtzite structure, which reduces the energy of the O-$2p_z$ orbitals by $\sim$0.1 eV with respect to the valence band maximum formed by O-$2p_{xy}$ orbitals. Therefore, in principle, the onset of direct band gap transitions for $E \bot c$ (O-$2p_{xy}$ $\rightarrow$ Zn-$3s$) and $E\parallel c$ (O-$2p_{z}$ $\rightarrow$ Zn-$3s$) should occur near 3.4 and 3.5 eV, respectively. Indeed, this band gap anisotropy has been reported by optical measurements \cite{Adachi1999}, but the measured energy difference is about 70 meV rather than the calculated 100 meV presented here. While good polarization selectivity is achieved at $d=6$ nm, the aloof-beam spectra are not fully polarized and contain mixed signals from both polarization directions. For the band gap onset area, the spectral identification is hindered by the signal contribution from the in-plane direction where the onset arises earlier than from the out-of-plane direction. However, as Fig.\ref{fig:BandGap} shows, we do observe some hint of anisotropy around the band gap region with a minor onset difference of 12 meV.

The key orientation-dependent transitions at 11.2 and 13.0 eV are identified as Zn-$3d$ $\rightarrow$ O-$2p_z$ near $\Gamma$ and Zn-$3d$ $\rightarrow$ O-$2p_{xy}$ near the boundaries of the hexagonal Brillouin zone (BZ), respectively. On the other hand, orientation-independent transitions at 9.1 and 13.5 eV are identified as Zn-$3d$ $\rightarrow$ O-$2p$ and O-$2p$ $\rightarrow$ Zn-$3s$, respectively. Our DFT calculations reveal the presence of another orientation-independent transition at 14.8 eV, which could not be distinctly observed in the recorded EELS spectra. It appears as a minor hump, primarily due to the presence of the nearby and broad surface plasmon. This transition is attributed to the electric dipole transition from occupied Zn-$3d$ to unoccupied O-$2p$ and occupied O-$2p$ to unoccupied \{Zn-$3d$, Zn-$3s$\} states. Some of the above-mentioned transitions occur at various $k$-points within the hexagonal BZ and between different sets of occupied and unoccupied bands. In Figs.~\ref{fig:appendix_A} and ~\ref{fig:appendix_B}, we made an attempt to mark some of these transitions having maximum optical-transition matrix elements using arrows.

Finally, we note that while localized surface plasmons (LSPs) are commonly observed in metallic nanorods \cite{Rossouw2013,Martin2014}, such distinct LSPs are not observed here due to the $\epsilon_1(\omega)$ behaviors of ZnO. Considering the vacuum/ZnO interface configuration studied here, surface plamons (SPs) are expected when $\epsilon_1(\omega)\leq-1$. However, as Fig. \ref{fig:dielectric_Adachi} in the Appendix shows, unlike in a metal, neither $\epsilon_{1,\bot}$ nor $\epsilon_{1,\parallel}$ of ZnO goes much below $-1$, and the energy range that supports SPs in ZnO is rather narrow, roughly from 15 to 17 eV. While there must be signals contributed from LSPs related to the nanorod geometry, the LSPs in the ZnO nanorod are suppressed due to its dielectric function, and the energy range where the LSPs are expected to reside does not affect the interband transition discussion presented here. While not shown here, we also performed plasmonic simulations of a 1100 by 40 nm ZnO nanorod using MNPBEM \cite{Hohenester2012} without observing the distinct LSPs that are typically observed in a metal nanorod.
 
\section{Summary}
\label{sec:summary}
We have characterized the electronic structures of single crystal w-ZnO nanorods by both intersecting and aloof beam configurations with a focused electron beam ($\alpha$ = 30 mrad) and moderate collection angle ($\beta$ = 16 mrad). The key anisotropic features of w-ZnO that are specific to the in-plane and out-of-plane directions were identified by DFT calculations. It was demonstrated experimentally that when keeping the beam-to-specimen orientation fixed, good polarization selectivity can only be achieved by the aloof beam configuration, not the intersecting one. We found that the good polarization selectivity brought by the aloof beam can be explained by analyzing the perpendicular component of the electric fields (Eq.~\ref{eq.E_perp}) generated by the fast electron used in the STEM-EELS experiments. The fact that the field directionality can be tuned by varying the impact parameter can be very useful in reducing spectral complexity and characterizing anisotropic nanomaterials such as the w-ZnO nanorods presented here.   

\section*{Acknowledgements}
P.E.B. acknowledges the financial supports of U.S. Department of Energy, Office of Science, Basic Energy Sciences under award no. DE-SC0005132.
S.S. acknowledges the support from the Office of Naval Research (ONR) grant N00014-21-1-2107.
D.V. acknowledges the support from National Science Foundation grant DMR-1954856. 
We also thank N. Dellby and M. Hoffman for discussions regarding the optical configuration of the microscope.
First-principles calculations were performed using the computational resources provided by the Rutgers University Parallel Computing clusters.

\section{Appendix}
\begin{figure}[htb!]
    \centering
    \includegraphics[width=0.48\textwidth]{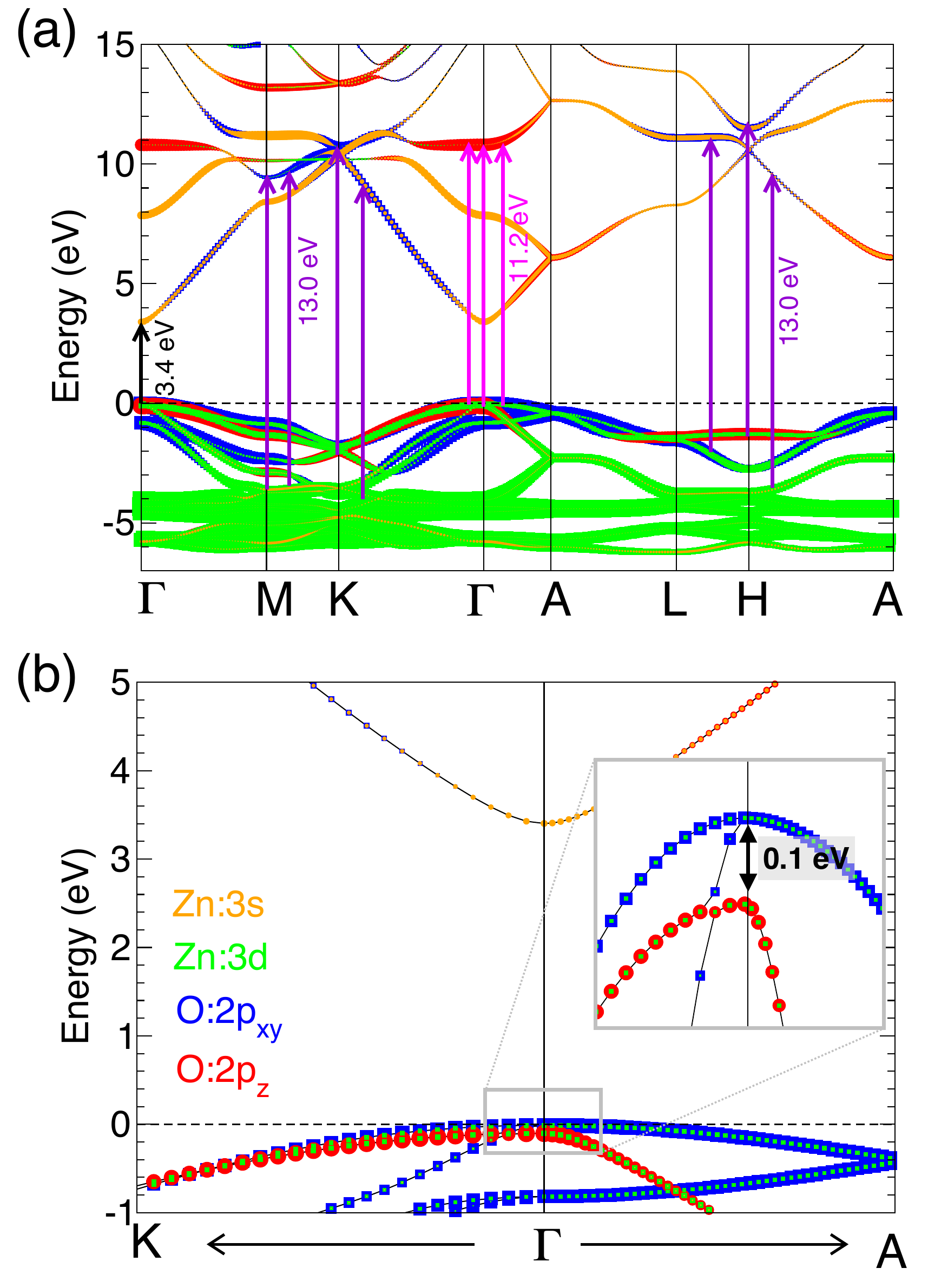}
    \caption{(a) Orientation-dependent interband transitions at 3.4, 11.2, and 13.0 eV marked using arrows in the orbital-resolved electronic band structure of w-ZnO. (b) An enlarged of bands near the Fermi level at zone center $\Gamma$. Inset shows a further amplification of valence band maximum near $\Gamma$. Crystal-field induced splitting of O-$2p_{xy}$ and O-$2p_z$ orbitals can be noticed. } 
    \label{fig:appendix_A}
\end{figure}

\begin{figure}[htb!]
    \centering
    \includegraphics[width=0.48\textwidth]{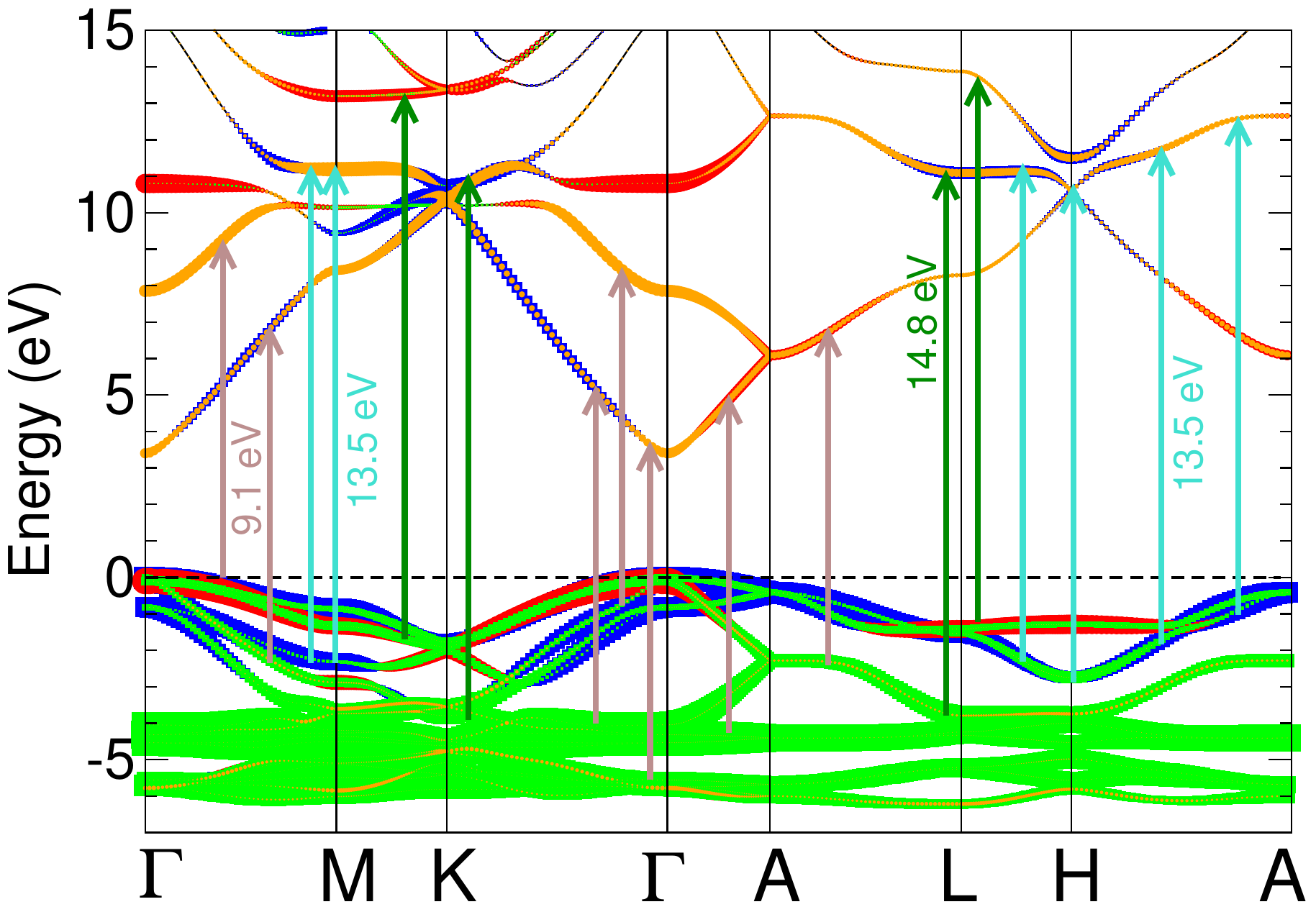}
    \caption{Orientation-independent interband transitions at 9.1, 13.5, and 14.8 eV marked using arrows in the orbital-resolved electronic band structure of w-ZnO. Only selected transitions having large optical-transition matrix elements are marked. } 
    \label{fig:appendix_B}
\end{figure}

Here we provide orbital-resolved electronic band structures marked with orientation-dependent (Fig.~\ref{fig:appendix_A}) and orientation-independent (Fig.~\ref{fig:appendix_B}) interband transitions. Only selected transitions having large optical-transition matrix elements are marked using arrows. There are numerous other transitions with relatively smaller transition probability which are not marked for the sake of clarity. Zn-$3s$ and Zn-$3p$ orbitals are depicted in orange and green colors, whereas O-$2p_{xy}$ and O-$2p_{z}$ orbitals are shown in blue and red colors, respectively, in all orbital-resolved band structures. 

\begin{figure}[htb!]
    \centering
    \includegraphics[width=0.48\textwidth]{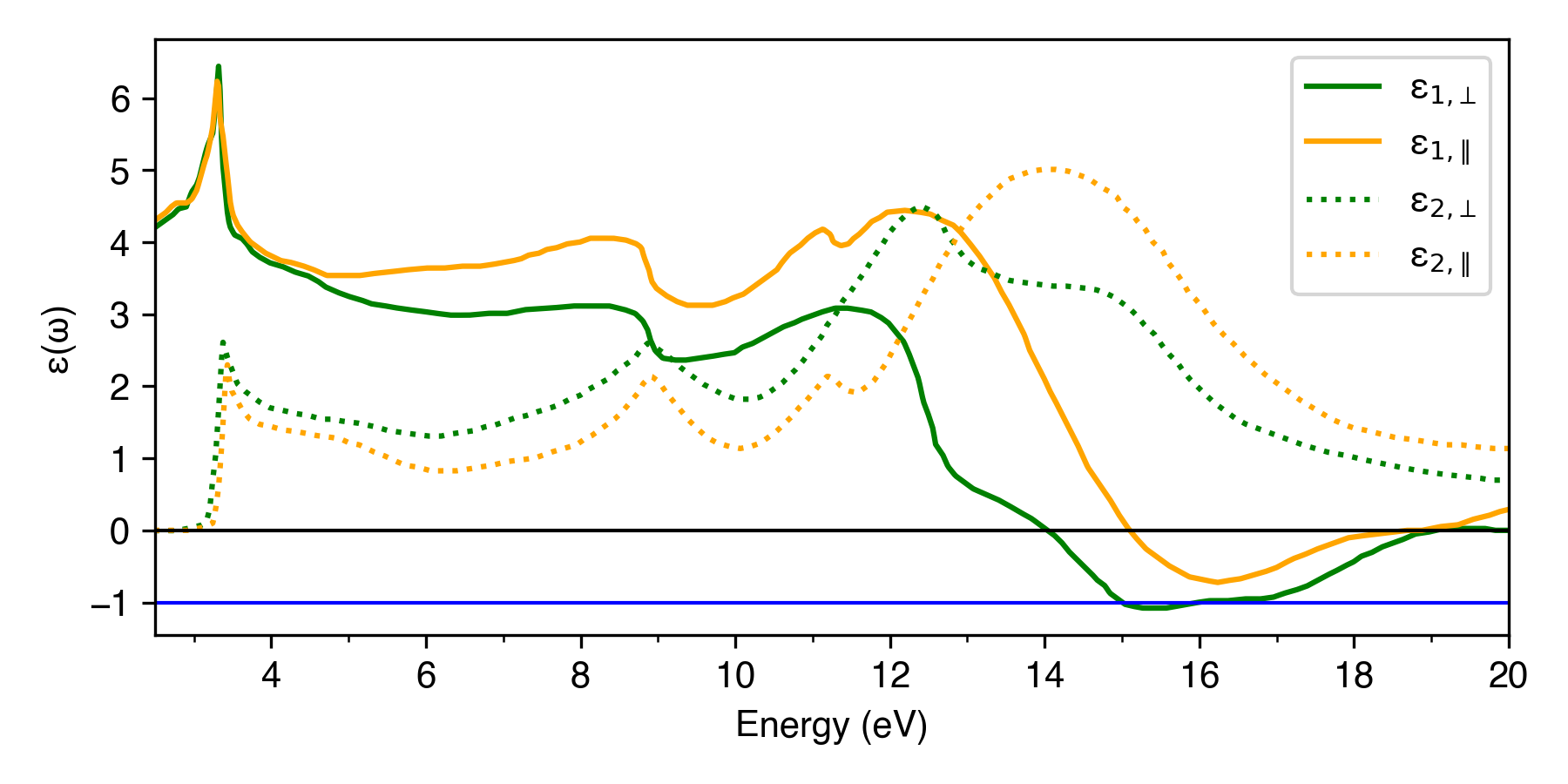}
    \caption{Replotted ZnO dielectric function using the data reported in Ref.~\cite{Adachi1999}. The real (solid curves) and imaginary (dotted curves) parts of the dielectric function along the in-plane and out-of-plane directions are shown in green and orange, respectively. Note that the energy range where $\epsilon_{1}<-1$ is between 15 and 17 eV.} 
    \label{fig:dielectric_Adachi}
\end{figure}

 \bibliography{references}

\end{document}